\documentclass[11pt,twoside]{article}

\usepackage{asp2006}
\usepackage{epsf}
\usepackage{psfig}
\usepackage{lscape}
\usepackage{natbib, graphicx}

\begin{document} 

\title{Supernova VLBI} %%% Fill in title
\author{Norbert Bartel}   %%% Fill in author names
\affil{York University, Toronto, ONT, M3J1P3, Canada} %%% Fill in author affiliations

\begin{abstract} %%% Abstract to run on from here.
We review VLBI observations of supernovae over the last quarter
century and discuss the prospect of imaging future supernovae with
space VLBI in the context of VSOP-2. From thousands of discovered
supernovae, most of them at cosmological distances, $\sim$50 have been
detected at radio wavelengths, most of them in relatively nearby
galaxies. All of the radio supernovae are Type II or Ib/c, which
originate from the explosion of massive progenitor stars. Of these, 12
were observed with VLBI and four of them, SN 1979C, SN 1986J, SN
1993J, and SN 1987A, could be imaged in detail, the former three with
VLBI.  In addition, supernovae or young supernova remnants were
discovered at radio wavelengths in highly dust-obscured galaxies, such
as M82, Arp 299, and Arp 220, and some of them could also be imaged in
detail. Four of the supernovae so far observed were sufficiently
bright to be detectable with VSOP-2.  With VSOP-2 the expansion of
supernovae can be monitored and investiated with unsurpassed angular
resolution, starting as early as the time of the supernova's
transition from its opaque to transparent stage.  Such studies can
reveal, in a movie, the aftermath of a supernova explosion shortly
after shock breakout.

%\keywords{supernovae:general, supernovae: individual (SN 1979C, SN 1980K, 
%SN 1986J, SN 1987A, SN 1993J, SN 1996cr, SN 2001em, SN 2001gd, SN 2003L, SN 2004et, 
%SN 2007gr, SN 2008D),  
%distance scale, instrumentation: interferometers, techniques: interferometric}

\end{abstract}

\section{Introduction} 
A supernova (SN), the explosion of a star, is one of the most
energetic single events in the universe.  Thousands of optical SNe are
now known but most of them are at cosmological distances.  Only
$\sim$50 have been detected at radio wavelengths and 12 observed with
VLBI, the most distant at almost 100 Mpc. In addition, several SNe and
supernova remnants (SNRs) were discovered at radio wavelengths in
dust-obscured galaxies, such as M82, Arp 299, and Arp 220, and
observed with VLBI. All of the SNe observed with VLBI are thought to
have resulted from the core collapse of a massive progenitor and emit
synchrotron radiation generated from the electrons accelerated in the
region where the shock front interacts with the circumstellar medium
(CSM) left over from the wind of the mass-losing progenitor.  In
addition, some of the radiation may come from the environment of the
stellar corpse, a neutron star or a black hole left over from the
explosion.  During the early stage of the expansion of a radio SN, the
radio lightcurve rises quickly because of the decreasing optical depth due to
synchrotron self-absorption or to thermal absorption in the ionized
CSM along the line of sight. It reaches its peak, first at high
frequencies later at lower frequencies, when the optical depth
decreases below unity, and it declines thereafter during the optically
thin stage of the SN while the SN is expanding
\citep[e.g.,][]{Chevalier1982a}.

For a typical shock front expansion velocity of 10,000 km s$^{-1}$,
the angular size of a SN at a distance of 4 Mpc expands at a
rate of 1 mas yr$^{-1}$. 
With
global VLBI and a wavelength of 4 cm, an angular resolution of
$\sim$0.6 mas can be obtained, allowing detailed investigations of the
expanding SN and the making of a movie. Global VLBI with VSOP-2
promises to increase the angular resolution at this wavelength to 0.2
mas and to a correspondingly higher resolution at shorter wavelengths, enabling
us to witness the aftermath of the explosion shortly after shock
breakout.

\section{Why supernova VLBI?}
The high angular resolution of VLBI affords us the following possibilities:
\begin{trivlist}
%\begin{enumerate}

\item{1}. The determination of the morphology.  A shell morphology
indicates the interaction of the ejecta with the CSM.  A
centrally-condensed morphology can indicate the presence of a central
source, such as a pulsar wind nebula or a black hole accretion disk system.  A
disk-like morphology suggests an optically thick object, as is
expected in the early stages of evolution.
\item{2.} The measurement of the angular size, $\theta$, which can give us the average
expansion velocity. It helps unravel whether synchrotron self-absorption
played a dominant role in the early evolution of the SN, and 
could reveal any possible presence of a relativistically expanding afterglow of a
gamma ray burst (GRB).  
\item{3.} The measurement of the expansion as a function of time, $t$, since 
shock breakout, parametrized by $\theta(t)\propto t^{m(t)}$, which can give us 
the SN's age and deceleration parameter, $m(t)$. 
\item{4.} Monitoring the expansion of the SN, which provides information 
on the density of the ejecta
and the CSM, and on the magnetic field, as a function of radius. Obtaining the density
profile of the CSM is like a ``time machine'' that reveals the last thousands of years
of the mass-loss history before the star died.
\item{5.} A geometric determination of the distance to the SN and its host galaxy 
using the ESM (expanding shock front method).
\item{6.} The making of a movie showing the evolution of the SN.

\end{trivlist}

\section{Supernovae observed with VLBI}
Supernova VLBI started with observations of SN 1979C in 1982
\citep{Bartel1985, Bartel+1985} shortly after the new, sensitive,
MKIII VLBI data acquisition system became available. Since then 11 more
optically identified SNe have been observed with VLBI
(Table \ref{tab1}). 
Eight are Type II, whose
progenitors are massive supergiants. Four are Type Ib/c, whose
progenitors are believed to be the cores of massive stars which have
lost their envelopes. A few Type Ib/c SNe were found to be associated with
GRBs. The sources in M82, Arp 299, and Arp 220 are not
optically identified but are likely also SNe of Type II or Ib/c or SNRs.
The distances range from 50 kpc for SN 1987A to almost 100 Mpc for SN
2003L. Their peak flux densities at 8 GHz reached values of 100 mJy and more, corresponding 
to very high brightness temperatures. 
Such SNe are examples for future targets of VLBI with VSOP-2. 

About 80\% or more of all VLBI observations focussed on just three SNe:
SN 1979C, SN 1986J, and SN 1993J. They were sufficiently
bright and extended that detailed images could be obtained and, for
the latter two, movies made.

\begin{table}[!ht]
  \begin{center}
  \caption{Characteristics of supernovae observed with VLBI}
  \label{tab1}
 {\scriptsize
  \begin{tabular}{lcccrllc}\noalign{\smallskip}\tableline
\noalign{\smallskip}
Name & Type & Host Galaxy &  Distance & $S_{8GHz}^1$ & 
$t_{peak}^2$ & $\theta_{peak}^3$ & Reference$^4$ \\ 
   &  &  &  (Mpc) & (mJy) & (yr) & (mas) & \\
\noalign{\smallskip}
\tableline
\noalign{\smallskip}

SN 1979C   & II & M100         & 15  & 7   & 0.9  & \,\,\,\,\,0.30  & 1 \\
SN 1980K   & II & NGC 6946     & 6   & 2   & 0.27 & $<$0.21 & 2 \\
SN 1986J   & II & NGC 891      & 10  & 100 & 2.7  & \,\,\,\,\,1.6  & 3 \\
SN 1987A   & II & LMC         & 0.05 & 80  & 0.005 & \,\,\,\,\,1.5  & 4 \\
SN 1993J   & II & M81          & 4   & 100  & 0.27  & \,\,\,\,\,0.55 & 5 \\
SN 1996cr  & II & Circinus Gal. & 4  & 150  & 2.7  & \,\,\,\,\,2.8    & 6 \\
SN 2001em  & Ib/c & NGC 7112     & 80 &  2  &  2   & \,\,\,\,\,0.11   & 7 \\
SN 2001gd  & II & NGC 5033       & 13 &   4 & 0.5  & \,\,\,\,\,0.16 & 8 \\
SN 2003L   & Ib/c & NGC 3506     & 92 &  3  &  0.2 & \,\,\,\,\,0.01 & 9 \\
SN 2004et  & II & NGC 6946       & 6  & 2   & 0.13 & $>$0.14 & 10 \\
SN 2007gr  & Ib/c & NGC 1058     & 10 &  $\raisebox{-0.3em}{$\stackrel{\textstyle <}{\sim}$}1$ & ? & \,\,\,\,\,? & 11 \\
SN 2008D  & Ib/c & NGC 2770      & 27  &  3  & 0.04 & \,\,\,\,\,0.01 & 12 \\
SN/SNR     & ?  & M82           & 4   &  ?  & ?    & \,\,\,\,\,?  & 13 \\
SN/SNR     & ?  & Arp 299        & 40 &  ?  & ?    & \,\,\,\,\,?  & 14  \\
SN/SNR     & ?  & Arp 220        & 77 &  ?  & ?    & \,\,\,\,\,?  & 15 \\
\noalign{\smallskip}

  \end{tabular}
}
 \end{center}

%\vspace{1mm} 
\scriptsize{ {\it Notes:}\\ 

$^1$ The approximate peak flux density of the radio lightcurve
measured at 8 GHz or extrapolated from lower frequencies.\\ 
$^2$ The
 approximate time after shock breakout when the 8-GHz radio lightcurve
 peaked, i.e., when the optical depth decreased to below unity. If
 $t_{peak}$ was not measured at 8 GHz, it is assumed to be $0.7\times
 t_{peak}$ at 5 GHz.  For items 1 and 2, see references (last column)
 and/or K. Weiler's home page at
 http://rsd-www.nrl.navy.mil/7213/weiler/sne-home.html.\\ 
$^3$ The
 approximate size of the SN at time $t_{peak}$ extrapolated from the
 size and expansion given in the references. For SN 1987A, an
 expansion velocity of 35,000~kms$^{-1}$\ is taken
 \citep{Gaensler+2007}. For the Type Ib/c SNe, for which only upper
 limits of sizes and expansions were available, an expansion velocity
 of 10,000~kms$^{-1}$\ is assumed.\\ 
$^4$ Latest, or latest
 comprehensive publications: 1) \citet{SN79C}; 2) \citet{Bartel1985};
 3) \citet{SN86J-Sci}; 4) \citet{Jauncey+1988}; 5) \citet{SN93J-Sci};
 6) \citet{Bauer+2008}; 7) \citet{SN2001em-2}; 8) \citet{Perez+2005};
 9) \citet{Soderberg+2005}; 10) \citet{Marti+2007}; 11) \citet{Paragi+2007}; 12)
 \citet{Soderberg+2008}; 13) \citet{Beswick+2006}; 14)
 \citet{NeffUT2004}; 15) \citet{Lonsdale+2006}. Also, see these
 references or references in there for the characteristics in the
 table.}

\end{table}

In the remainder we will first focus on these three SNe. Then we will
discuss the other optically identified SNe and the unidentified
SNe/SNRs and lastly address prospects for space VLBI in the context of
VSOP-2.

\section{Supernovae with detailed VLBI images}

\subsection{SN 1993J}
\begin{figure}[!ht] 
% \vspace*{-2.0 cm}
\begin{center}
\includegraphics[width=5.3in]{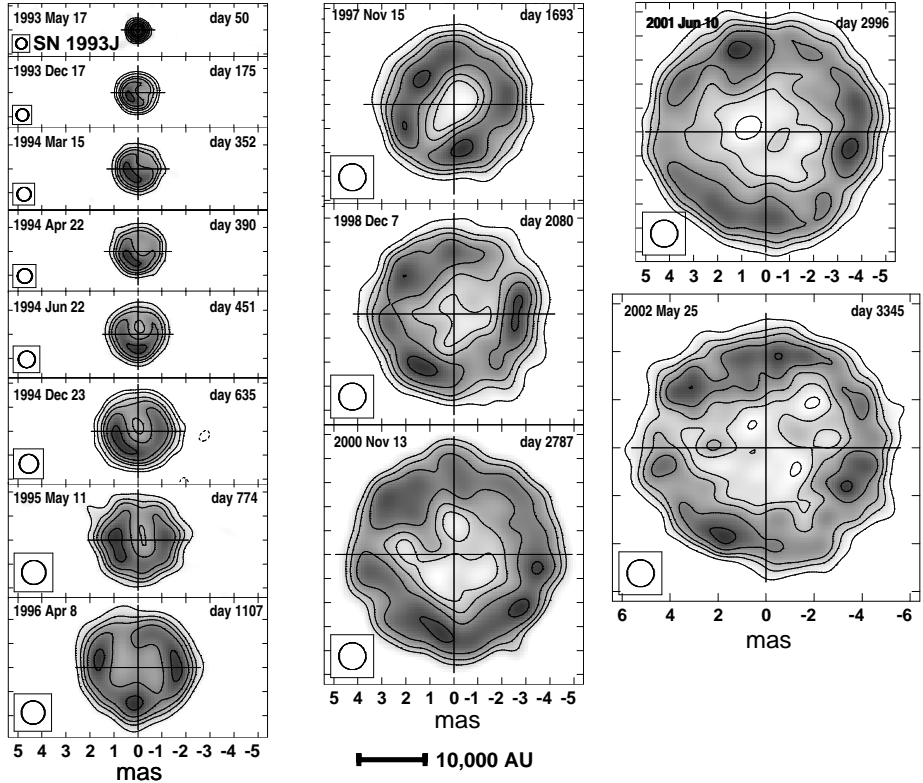} 
% \vspace*{-1.0 cm}
 \caption{A sequence of 8.4-GHz VLBI images of SN 1993J from $t=50$ to
3345~d after shock breakout \citep{SN93J-3}. The contours are at -1,
1, 2, 4, \dots, 32, 45, 64, and 90\% of the peak brightness. For images
here and hereafter, the scale on the vertical axis is the same as that
on the horizontal axis, the restoring beam is plotted in the lower
left, and north is up and east to the left. (To download SN VLBI
movies, please visit: http://www.yorku.ca/bartel.)}
   \label{fig1}
\end{center}
\end{figure}

SN 1993J in M81 is, after SN 1987A, the most intensely observed SN ever. It
provides an exemplary case to witness, with increasing relative
angular resolution, the evolution of a SN over time. VLBI observations
started $\sim$ 30~d after shock breakout \citep{Marcaide+1994,
SN93J-Nat} and showed that the SN expanded almost freely early on
\citep{SN93J-Nat}.  A first detailed image was obtained only several
months later after the SN had expanded sufficiently \citep[see also,
Bartel et al. 1995]{Marcaide+1995a}\nocite{BartelBR1995}.
Fig.\,\ref{fig1} shows an example of a sequence of images of
the expanding shell of the SN \citep[see Marcaide et al. 1995b for an
earlier sequence]{SN93J-3}\nocite{Marcaide+1995b}.

The explosion occurred at a location pinpointed with 160~AU accuracy
in the galactic reference frame of the host galaxy M81.  From there,
the SN expanded isotropically to within 5.5$\%$ \citep{SN93J-1}.

Initially, the SN expanded rapidly with a velocity of $\sim$
20,000~kms$^{-1}$. Then it was found that the expansion decelerated slightly
\citep{Marcaide+1997} and that the deceleration changed over time,
$t$, \citep[Fig. 2, left panel;][2002]{SN93J-Sci} with the velocity decreasing to
$\sim$10,000~kms$^{-1}$.
These measurements are fairly consistent with a hydrodynamic model
\citep{MioduszewskiDB2001} and provide insight into the physical
interplay between the ejecta and the CSM, and the mass-loss history of
the progenitor \citep{SN93J-2}.

\begin{figure}[ht]
% \vspace*{-2.0 cm}
\begin{center}
 \includegraphics[width=5.4in]{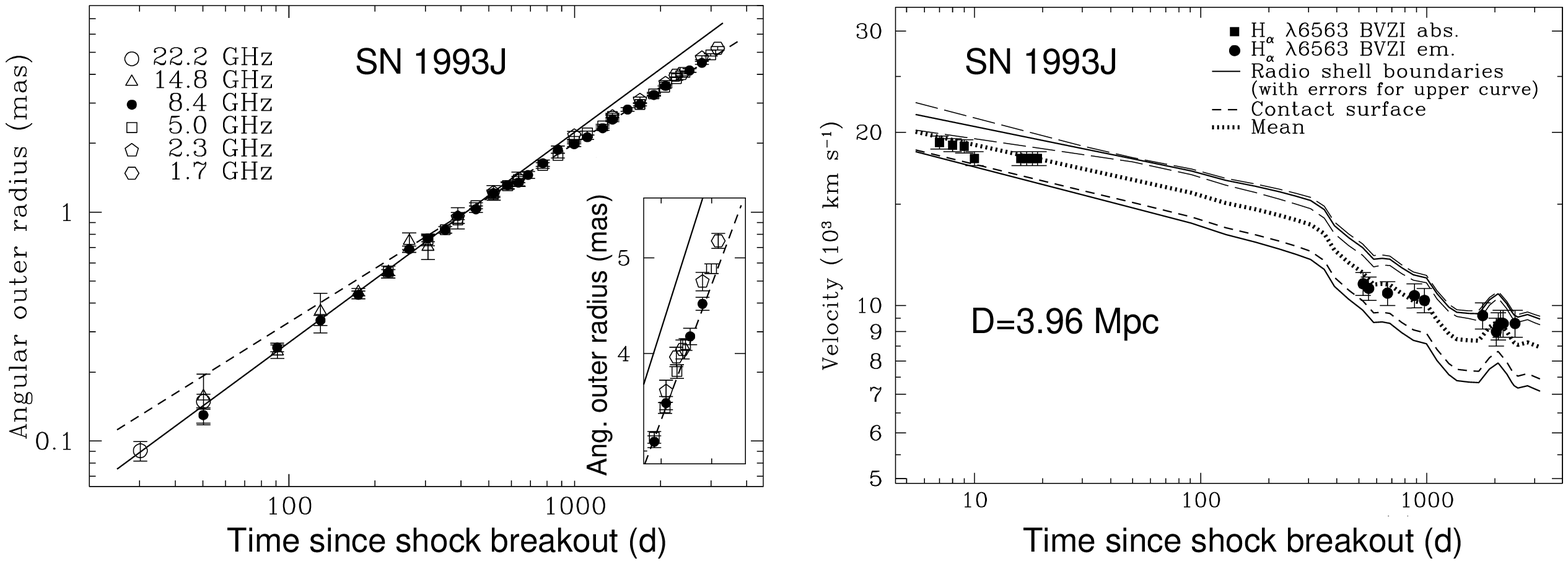}
% \vspace*{-1.0 cm}
 \caption{Left: The expansion of SN 1993J.  The straight lines give
fits which show the changing deceleration of the expansion with
$\theta\propto t^{0.919\pm0.019}$ for the solid line and $\theta\propto
t^{0.781\pm0.009}$ for the dashed line.  The inset gives a zoomed-in
version of the latest part of the expansion curve \citep{SN93J-2}.  
Right: The velocity of the forward and reverse
shocks (outer and inner radio shell) with their mean fit to the
maximum H$\alpha$ line velocities, giving the distance to SN 1993J and
M81 as indicated in the figure \citep{SN93J-4}.}
   \label{fig2}
\end{center}
\end{figure}

SN 1993J provided also the best example of geometrically determining
a distance using the expanding shock front method
(ESM). The distance is derived to be 3.96$\pm$0.29~Mpc
(Fig.\,\ref{fig2}, right panel), which, when compared with Cepheid distances, is
somewhat larger than \citet{Freedman+1994}'s value of
3.63$\pm$0.34~Mpc but comparable to \citet{Huterer+1995}'s value of
3.93$\pm$0.26~Mpc.

\subsection{SN 1986J}
\begin{figure}[!ht]
% \vspace*{-2.0 cm}
\begin{center}
 \includegraphics[width=5.4in]{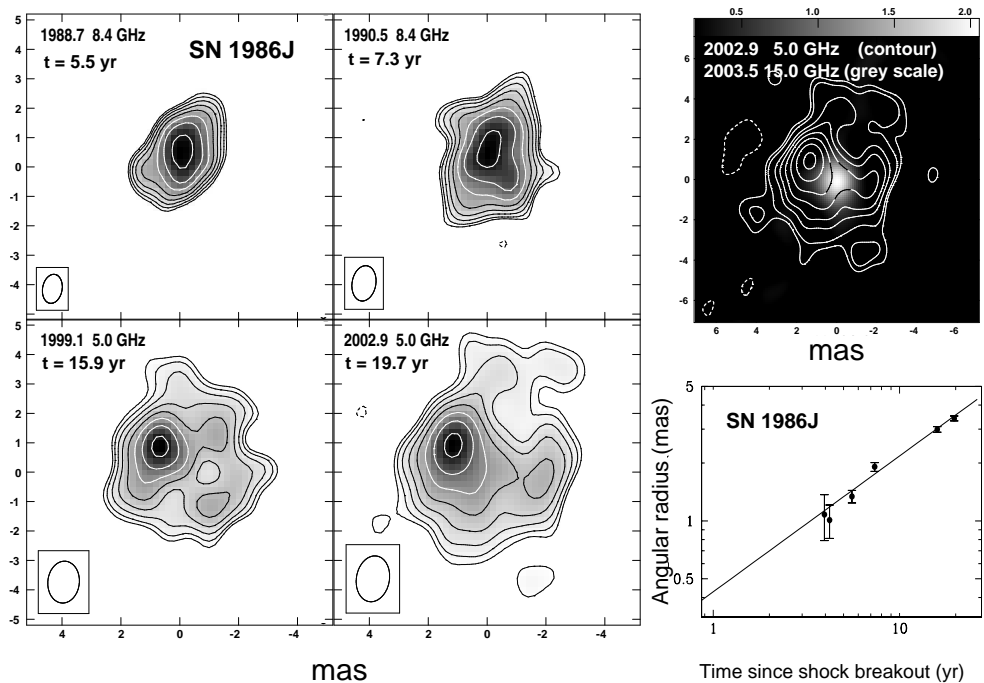}
% \vspace*{-1.0 cm}
 \caption{Left: A series of VLBI images of SN 1986J from 1988.7 to 2002.9 with
observing frequency and time after shock
breakout (1983.2) indicated. The contours are at $-$8, 8, 11.3, 16, 22.6, 32, \dots, and 90.5\%
of the peak brightness. Top right: A 15-GHz image from 2003.5 showing
the central component in grey scale and the 5-GHz image from 2002.9 in (slightly different)
contours, in the same reference frame. Bottom right: The expansion of
SN 1986J with a best fit of $\theta\propto t^{0.71\pm0.11}$
\citep{SN86J-Cospar}.}
   \label{fig3}
\end{center}
\end{figure}

SN 1986J in NGC 891 is an intriguing extragalactic SN, the first one for which
a compact component was found to emerge in the projected center of the expanding
shell, $\sim$20 years after the explosion. It was also the first
optically identified SN for which a detailed image could be obtained
\citep{Bartel+1991}. A series of images is shown in Fig.\,\ref{fig3}.
The new component has a spectrum which was inverted between 5 and 20 GHz
when first discovered. The peak is slowly moving to lower
frequencies. The inverted part of the spectrum can be interpreted as
being due to free-free absorption within the SN shell with the
absorption weakening as the shell expands.  The new component could be
radio emission associated with accretion onto a black hole or the
nebula formed around an energetic young neutron star in the center of
SN~1986J, which would be the first direct link of a black hole or a
neutron star to a modern SN. Its spectral luminosity between 14
and 43 GHz is $\sim200$ times that of the Crab Nebula, which is in the range
expected for young pulsar wind nebulae \citep{BandieraPS1984}. However, since the new
component is in the center only in projection, it could also be caused
by the interaction of the shock with a particularly dense CSM
condensation at the far or near end of the shell. The expansion of the
SN is moderately decelerated (see, Fig \,\ref{fig3}, lower right panel).
%\citep{SN86J-1, SN86J-Cospar}.

\subsection{SN 1979C}
SN 1979C in M100 in Virgo has been studied since 1982, and because of its distance of
$\sim$15~Mpc could only recently be resolved well enough to reveal its
shell-like structure \citep[Fig. 4, left panel;][]{SN79C-2}. The most interesting
characteristic is that its expansion is consistent with being almost
free up to $t\sim$20 yr (Fig. 4, right panel). However, significant deceleration may start around
this time. From fits to the radio lightcurve, a relatively high
progenitor mass-loss rate of 1 to
$2\times10^{-4}$~M\raisebox{-.6ex}{$\odot$}~yr$^{-1}$ for a wind
velocity of 10~kms$^{-1}$\ was estimated \citep{Weiler+1991,
Montes+2000}. However, the kinetic energy of the SN explosion would
have to be $\sim2\times10^{52}$~erg to sweep up the mass that
corresponds to such mass-loss, more than an order of magnitude larger
than is thought appropriate.  The mass-loss rate is therefore likely
an order of magnitude smaller.  Estimates of mass-loss to
wind-velocity ratios from radio lightcurve fittings may therefore have
to be interpreted with caution \citep{SN79C}.

\begin{figure}[!ht]
% \vspace*{-2.0 cm}
\begin{center}
 \includegraphics[width=4.9in]{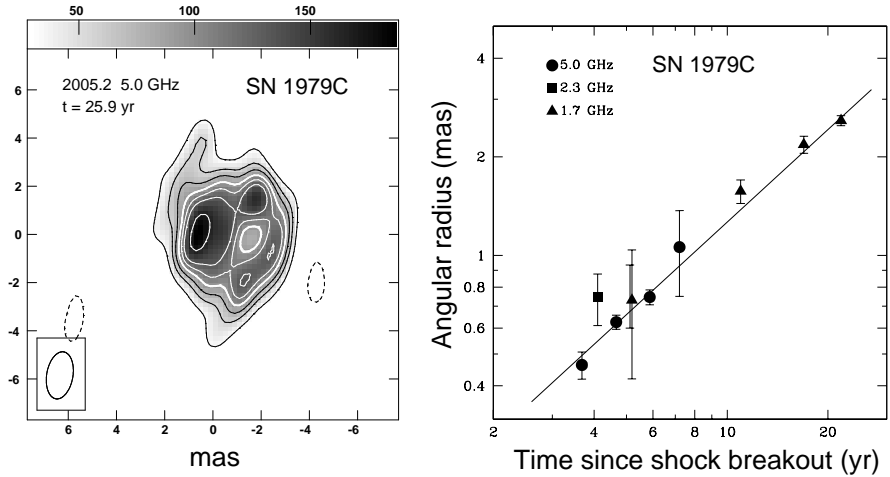}
% \vspace*{-1.0 cm}
 \caption{Left: A 5.0-GHz VLBI image of SN 1979C at $t=25.9$~yr. The contour
levels are at $-17$, 17, 30, 40, {\bf 50} (emphasized), 60, 70, and 90\% of the
peak
brightness of 186~$\mu$Jy bm$^{-1}$. The grey scale, in $\mu$Jy bm$^{-1}$,
is given at the top \citep{SN79C-2}. Right: The expansion of SN
1979C with a best fit of $\theta\propto t^{0.94\pm0.03}$ for a shell 
model \citep{SN79C}.}
\label{fig4}
\end{center}
\end{figure}

\section{Other optically identified supernovae}

The only other optically identified SN for which detailed
images could be obtained is SN 1987A in the LMC. The images were not
obtained with VLBI, but rather with the Australia Telescope Compact
Array (ATCA).  SN 1987A is by far the closest SN that has ever been
observed with VLBI. Its radio lightcurve peaked at $\sim$80 mJy, but
when VLBI observations were made in a heroic effort $t=5$ d after
shock breakout, the SN was already completely resolved. Nevertheless,
a lower limit on the expansion velocity of 19,000 km$^{-1}$ could be
derived \citep{Jauncey+1988}. In fact, the SN expanded with a velocity
of $\raisebox{-0.3em}{$\stackrel{\textstyle >}{\sim}$}$35,000~kms$^{-1}$\ 
and slowed down to 3600 to 5200~kms$^{-1}$\ for
$t>1800$~d, consistent with the progenitor being a blue supergiant. 
SN~1987A has a fairly round shell or ring-like morphology with a
modulated brightness around the ridge \citep{Manchester+2002,
Gaensler+2007}.

For the other optically identified SNe, no detailed images could be
obtained.  Four of them are Type II. SN 1980K and SN 2004et in NGC
6946 were observed only once with VLBI. The angular size of SN 1980K
was determined to be $\raisebox{-0.3em}{$\stackrel{\textstyle <}{\sim}$}2.0$ mas for a shell model
\citep{Bartel1985} corresponding to an average expansion velocity of
$\raisebox{-0.3em}{$\stackrel{\textstyle <}{\sim}$}$11,000~kms$^{-1}$. For SN 2004et a marginally resolved image, with
an indication of an asymmetric brightness distribution, and a lower
expansion velocity limit of 15,700$\pm$2000~kms$^{-1}$\ was obtained, and it was
inferred that synchrotron self-absorption is not relevant for this SN
\citep{Marti+2007}. For SN~2001gd a size of 0.37$\pm$0.08~mas was
obtained a couple of years after the explosion
\citep{Perez+2005}. Similarly, for SN 1996cr a preliminary size of 10~mas 
was determined, which helped in the interpretation of this
SN as possibly being related to SN~1987A \citep{Bauer+2008}.

The remaining four are Type Ib/c SNe.  For SN 2007gr, only a detection
with VLBI was reported \citep{Paragi+2007}.  For the others, SN
2001em, SN 2003L, and SN 2008D, upper limits on sizes with
corresponding upper limits on average expansion velocities were
obtained. For the latter two SNe, these upper limits are relativistic
\citep{Soderberg+2005, Soderberg+2008}. For SN 2003L, at 92 Mpc the
most distant of the SNe in Table 1, the upper limit helped to
distinguish between absorption models, suggesting a (model-dependent)
non-relativistic nature of the SN \citep{Soderberg+2005}. For SN
2001em, a non-relativistic limit of the expansion velocity could be
determined directly.  The SN was considered to possibly be related to
a GRB event but with a misaligned relativistic jet
\citep{GranotR2004}.  However, VLBI measurements of an expansion
velocity of only 5800$\pm$10,000~kms$^{-1}$\ and a proper motion
velocity of only 33,000$\pm$34,000~kms$^{-1}$\ disfavor this
possibility \citep{SN2001em-2}.

\section{Optically unidentified supernovae and supernova remnants}
There are a few dust-obscured starburst galaxies with compact radio
sources that were found to be SNe or SNRs: M82 \citep{Beswick+2006},
Arp 299 \citep{NeffUT2004}, Arp 220 \citep{Lonsdale+2006}. In two
other galaxies, NGC 6240 \citep{GallimoreB2004}, and Mrk 273
\citep{Bondi+2005}, compact radio sources were found with VLBI, which
are also thought to be possibly SNe or SNRs.  The most and longest
studied are those in M82. Two sources could be imaged at several
epochs. The weaker one, 43.31+592, was found to be shell-type. It
expands with a velocity of 9000-11,000~kms$^{-1}$, which is typical for
SNe. The strongest of all compact sources in M82, 41.95+575, however,
is puzzling. The earliest clearly resolved images suggested a
shell-like structure, albeit quite elongated with two components at
opposite ends \citep{Bartel+1987, WilkinsondB1990}. Recently, an
expansion velocity of 1500-2000~kms$^{-1}$\ was determined, which is small for
a SN. In addition, the bipolar appearance became stronger, and
the nature of this source is now less clear \citep{Beswick+2006}.

\section{Supernova space VLBI in the context of VSOP-2}
From the optically identified SNe studied so far with VLBI, four had
flux densities peaking at $\sim$100 mJy when they became optically
thin: SN 1986J, SN 1987A, SN 1993J, SN 1996cr.  Their sizes
ranged from 0.55 to 2.8 mas at these times (see Table \ref{tab1}).  VLBI observations with
VSOP-2 would allow making detailed images of such SNe starting at the
earliest possible time when the optical depth decreases below unity
and the SN becomes optically thin.  ``Filming'' in radio the earliest
stages of a SN evolution promises to add invaluable information to our
understanding of the aftermath of the explosion.

%\bibliographystyle{apj}
%\bibliography{/home/bartel/paper/nb-bib}

\end{document}